\documentclass[prd,preprintnumbers,12pt]{revtex4}
\pdfoutput=1
\usepackage{epsfig}
\usepackage{graphicx}
\usepackage{bm}
\usepackage{amsmath}
\usepackage{amsfonts}
\usepackage{amssymb}

\newcommand{\be}{\begin{equation}}
\newcommand{\dd}{\displaystyle}
\newcommand{\ee}{\end{equation}}
\newcommand{\bea}{\begin{eqnarray}}
\newcommand{\eea}{\end{eqnarray}}

\newcommand{\nn}{\nonumber}

\def\nn{\nonumber}

 \def\slash#1{\setbox0=\hbox{$#1$}#1\hskip-\wd0\dimen0=5pt\advance
       \dimen0 by-\ht0\advance\dimen0 by\dp0\lower0.5\dimen0\hbox
         to\wd0{\hss\sl/\/\hss}}
\def\be{\begin{equation}}
\def\dd{\displaystyle}
\def\ee{\end{equation}}

\def\bea{\begin{eqnarray}}
\def\eea{\end{eqnarray}}

\def\7{\tilde}
\def\8{\hat}

 \def\slash#1{\setbox0=\hbox{$#1$}#1\hskip-\wd0\dimen0=5pt\advance
       \dimen0 by-\ht0\advance\dimen0 by\dp0\lower0.5\dimen0\hbox
         to\wd0{\hss\sl/\/\hss}}

\def\tZ{\tilde Z}

\def\tP{\tilde P}
\def\tB{\tilde B}

\def\tZ{\tilde Z}
\def\tX{\tilde X}
\def\tM{\tilde M}

\def\cG{{\cal G}}
\def\cM{{\cal M}}
\def\cP{{\cal P}}
\usepackage{color}
\usepackage{color}

\newcommand{\eq}[1]{(\ref{eq:#1})}

\begin{document}

{\hfill {\rm ICCUB-19-016}}
\title{
Non-relativistic $k$-contractions of the Coadjoint Poincar\'e algebra }

\author{Andrea Barducci}\email{barducci@fi.infn.it} \affiliation{Department of Physics, University of Florence and INFN Via G. Sansone 1, 50019 Sesto Fiorentino (FI), Italy}
\author{Roberto Casalbuoni}\email{casalbuoni@fi.infn.it}
\affiliation{Department of Physics, University of Florence and INFN Via G. Sansone 1, 50019 Sesto Fiorentino (FI), Italy}
\author{Joaquim Gomis}\email{joaquim.gomis@ub.edu}\affiliation{Departament de F\'isica Qu\`antica i Astrof\'isica\\ and
Institut de Ci\`encies del Cosmos (ICCUB), Universitat de Barcelona\\ Mart\'i i Franqu\`es , ES-08028 Barcelona, Spain
}


\begin{abstract}
 We study a class of extensions of the k-contracted Poincar\'e algebra under the hypothesis
of generalising the Bargmann algebra and its central charge. 
As we will see this type of contractions will lead in a natural way to consider
the codajoint Poincar\'e algebra and some of their contractions.
 Among them there is one such that considering the quotient of it by a suitable ideal, the (stringy) p-brane
Galilei algebra is recovered. 
 \end{abstract}
\maketitle

\section{Introduction}
Non-relativistic (NR) holography \cite{Sachdev,Liu} has played a central role in the
renewed interest to study Newton-Cartan gravity \cite{Cartan1,Cartan2} and  generalizations, 
\cite{Papageorgiou:2009zc,Bergshoeff:2017btm,Bergshoeff-Rosseel,Obers-Hartong,
Hansen:2018ofj,Hansen:2019vqf,Aviles:2018jzw,Ozdemir:2019orp}.
The NR symmetries play a crucial role in their construction. 
 Since the NR  limit is 
not unique \cite{Batlle:2016iel,Barducci:2018wuj} for extended objects, we could construct different NR gravities that  could couple these NR extended objects to gravity
\cite{Andringa:2012uz,Bergshoeff:2018vfn,Gomis:2019zyu,Aviles:2019xed,Gallegos:2019icg}.
Inspired by this fact we have introduced
$k$-contractions of Galilei  type of the Poincar\'e algebra \cite{Barducci:2018wuj}. In the case of Newton-Hooke algebras see
also \cite{Gomis:2005pg,Brugues:2006yd,Gomis:2019fdh}. 
Due to the  contraction procedure these galilean algebras have no extensions. 
Instead the symmetry algebras associated to NR p-branes have extensions  
\cite{Brugues:2006yd,  Brugues:2004an}


In this paper we look for all the possible 
extensions of the k-contracted Poincar\'e algebra that generalise  the 
Bargmann algebra \cite{Bargmann:1954gh}.  More concretely we look for extensions of these algebras that generalise the commutation relations 
\be
[\tB_{a}, \tP_0]=-i\tP_a,~~~[\tB_{a}, \tP_b]=- i\delta_{ab}\tZ\label{eq:8}
\ee
where $\tB_a$ are the boosts, $\tP_0$ and $\tP_a$ are the energy and the momentum, and $\tZ$ is the central charge.
As we will see this generalization is not possible by  introducing only a vector generator
$\tZ_\mu$ that generalises $\tZ$. We will need to introduce a generator  $\tZ_{\mu\nu}$ behaving as an antisymmetric tensor under the Lorentz group. In order words 
we need to double
the initial Poincar\'e generators. If we demand that the enlarged 
Lie  algebra has an invariant non degenerate metric, 
according to the double extension 
procedure \cite{medina,Figueroa-OFarrill:1995opp,Matulich:2019cdo},
the algebra is unique and it is the coadjoint algebra of the Poincar\'e algebra.
At this point we will introduce the k-NR contraction of this algebra. We will see that 
there are several possibilities, one of them contains an ideal such that if we consider a
quotient algebra  we will find the
(stringy) p-brane algebra \cite{Brugues:2004an,Brugues:2006yd}.

The organization of the paper is as follows. In section 2 we will 
review the k-contractions of  the Poincar\'e algebra without extensions. 
In section 3 we will
write the generalization to extended objects of the Bargmann algebra by extending the Poincar\'e algebra introducing
  a vector generator $Z_\mu$.
In section 4 we will prove that it is not possible to obtain the Bargmann algebra    introducing  $Z_\mu$ only.
In section 5 we will consider the k-contractions of  the coadjoint Poincar\'e algebra with
generators $P_\mu, M_{\mu\nu}, Z_\mu, Z_{\mu\nu}$.
Section 5 is devoted to conclusions. There is an appendix where certain aspects of the double extension procedure are reviewed.

\section{Description of the $k$-contractions for the Poincar\'e group}\label{sec:II}

In this Section we will review the  $k$-contractions of Galilei  type of the Poincar\'e algebra, as defined in \cite{Barducci:2018wuj}, see also \cite{Brugues:2004an,Gomis:2005pg,Brugues:2006yd}.  The generators of the Poincar\'e  algebra in $D+1$ space-time dimensions $M_{\mu\nu}$ and $P_\mu$, satisfy the commutation relations ,
\bea
 \left[M_{\mu\nu},M_{\rho\sigma}\right]&=&i(\eta_{\mu\rho}M_{\nu\sigma}+\eta_{\nu\sigma}M_{\mu\rho}
 -\eta_{\mu\sigma}M_{\nu\rho}
  -\eta_{\nu\rho}M_{\mu\sigma}),
\nonumber \\
  \left[M_{\mu\nu},P_{\rho}\right]&=&i(\eta_{\mu\rho}P_{\nu}-\eta_{\nu\rho}P_{\mu}),
\nn\\
  \left[P_{\mu},P_{\nu}\right]&=&0 
  \label{eq:1.1}
\eea
with $\eta_{\mu\nu}=(-;+,\cdots,+)$ and $\mu,\nu=0,1,...,D$.

To define the  $k$ contractions of  this  algebra we begin  by partitioning the $D+1$ dimensional space-time in a $k$ dimensional Minkowskian part and in a $D+1-k$ dimensional Euclidean one, 
   by introducing 
 the following set of labels for the space-time coordinates
\bea
&&\alpha,\beta =0,1,\cdots,k-1,~~~\eta_{\alpha\beta}=(-;+,\cdots,+),\nn\\
&&a,b =k,\cdots,D,~~~\eta_{ab}=(+,+,\cdots,+).
\eea

Then, let us  consider  the following two subgroups of $ISO(1,D)$: the Poincar\'e subgroup in $k$ dimensions, $ISO(1,k-1)$ and the euclidean group of toto-translations in $D+1-k$ dimensions, $ISO(D+1-k)$,generated respectively by
\be
ISO(1,k-1):~~~M_{\alpha\beta},~~P_\alpha,~~~\alpha,\beta =0,1,\cdots,k-1,
\ee
\be
ISO(D+1-k):~~~M_{ab},~~P_a,~~~a,b =k,\cdots,D.
\ee
In these notations the generators of $ISO(1,D)$ are
\be
ISO(1,D):~~~M_{\alpha\beta},~~~M_{ab},~~P_\alpha,~~~P_a,~~~M_{\alpha b}\equiv B_{\alpha b}.
\ee
 Note that the generalized boosts $ B_{\alpha b}$ inter-twin the momenta generators lying in the two subalgebras.
 
 In \cite{Barducci:2018wuj} we have considered  two types of contractions, Galilei and Carroll type. In this paper we will concentrate on the Galilei type only. These $k$-contractions   represent a generalization of   the Galilei algebra with no central extension, see \cite {Brugues:2004an,Gomis:2005pg, Brugues:2006yd}.

At the  Lie algebra level the contractions are made on the momenta and on the boosts as follows
\bea
\tilde P_a ={\dd \frac 1\omega}P_a,~~~
 \tB_{\alpha a}= {\dd \frac 1\omega} B_{\alpha a},\label{eq:7}
\eea       
and taking the limit $\omega\to\infty$.  The tilde generators  will be the ones associated to the "non-relativistic" algebra,  once the limit is taken, and the corresponding operators will be denoted as the "contracted" generators. 
 Apart from the Lorentz rotations and the spatial rotations the commutation relations of the boots  $ B_{\alpha b}$ are
\be
[\tilde B_{\alpha a},\tB_{\beta c}]=0,~~~[\tilde B_{\alpha a}, \tilde P_\beta] =i\eta_{\alpha\beta}\tP_a,~~~[\tilde B_{\alpha a}, \tilde P_b]=0,\label{eq:1.11}
 \ee

Since the Poincar\'e algebra is invariant under a global rescaling of the momenta, the previous definition of the contractions is equivalent to take $\tilde P_\alpha =\omega P_\alpha$, leaving unchanged the scaling of the boosts.

\section{The Galilei algebra with 
extensions}

The usual Bargmann algebra is generated by the energy $\tP_0$, spatial momenta
$\tP_a$, rotation generators $\tM_{ab}$ and boosts $\tB_a$, plus a central charge $\tZ$, corresponding to the mass. In previous notations: $\alpha=0$, $a=1,\cdots,D$. 
The relevant non-vanishing commutators are:
\be
[\tB_{a}, \tP_0]=-i\tP_a,~~~[\tB_{a}, \tP_b]=- i\eta_{ab}\tZ\label{eq:8}
\ee
with $\tZ$ commuting with all the generators and $\tB_{0a}\equiv \tB_a$, therefore
 $\tZ$ is a central charge.

The Bargmann algebra \cite{Bargmann:1954gh}
 cannot be obtained from  the  non-relativistic $k=1$-contraction of Poincar\'e. We should enlarge the starting algebra before contraction. One should consider  the direct product of Poincar\'e$\otimes U(1)$ \cite{Aldaya:1985plo}.
Then, the main point  
is to mix the translation generator $P_0$ with the $U(1)$ charge $Z$ and after
to do the limit
.

\be
P_0=\tP_0+\omega^{{\color{blue}2}}
\tZ,~~~Z=\tZ
\ee
maintaining the relations (\ref{eq:7}) for the other 
generators. From these relations one gets immediately the commutators of eq. \eq{8}
{ in the limit}.

 Now we want
 to generalise the eq. \eq{8} to the case of a generic $k$-contraction. In order to do that we make 
 the minimal ansatz by introducing the generators $\tZ_\alpha$ with commutation relations 
 \be
[\tB_{\alpha a}, \tP_\beta]=i\eta_{\alpha\beta}\tP_a,~~~[\tB_{\alpha a}, \tP_b]=- i\eta_{ab}\tZ_\alpha\label{eq:9}
\ee
%
%
%
%
%
%
 Notice that we are not doing any assumption about the boost commutators.
In order to preserve the covariance in the Minkowski subspace, $\tZ_\alpha$ should transform as a Lorentz vector in $k$ dimensions. However, since we want to obtain the algebra \eq{9} as a $k$-contraction of a relativistic algebra, we need to extend the Poincar\'e group introducing a $D+1$ dimensional Lorentz vector, $Z_\mu$.  

Since the vector space spanned by the charges $Z_\mu$ is isomorphic to the one spanned by the momenta, the Poincar\'e algebra can be extended 
with   a method inspired by the double extension procedure \cite{medina} 
\cite{Figueroa-OFarrill:1995opp}, see also \cite{Matulich:2019cdo},
that works for a Lie algebra and its dual, without requirement about the existence or not of a non-degenerate metric. According to the Appendix
 we define  a
 mapping "star" among the translations generators $P_\mu$ 
 and their dual generators living in the dual vector space,  $P_{\mu}^*=Z_{\mu}$. 
We have
\bea
  \left[M_{\mu\nu},Z_{\rho}\right]&=& \left[M_{\mu\nu},P_{\rho}^*\right]=i(\eta_{\mu\rho}P_{\nu}^*-\eta_{\nu\rho}P_{\mu}^*),
=i(\eta_{\mu\rho}Z_{\nu}-\eta_{\nu\rho}Z_{\mu}),
\nn\\
   \left[Z_{\mu},Z_{\nu}\right]&=& \left[P_{\mu}^*,P_{\nu}^*\right]=0 ,~~~  \left[Z_{\mu},P_{\nu}\right]=\left[P_{\mu}^*,P_{\nu}\right]=\left[P_{\mu},P_{\nu}\right]^*=0,  \label{eq:10}
\eea
notice that the $Z_\mu$'s commute among themselves.

In the next Section we will show that no contraction of the Poincar\'e algebra extended with the vector charges $Z_\mu$ is able to reproduce the commutators \eq{8}.
 
 \section{A no-go theorem}
 Since both $P_\mu$ and  $Z_\mu$ satisfy analogous commutation relations 
  we introduce the vectors
\be
X_\alpha=\left(\begin{array}{c}P_\alpha \\
Z_\alpha\end{array}\right),~~~X_a=\left(\begin{array}{c}P_a \\
Z_a\end{array}\right) \label{eq:23}
\ee
The most general contraction preserving the covariance in the Minkowskian and in
the  Euclidean subspaces can be expressed in terms of two matrices, $A$  and $B$:
\be
X_\alpha = A \tX_\alpha,~~~X_a = B \tX_a
\ee
where $A, \,B$ are $2\times 2$ matrices.

We have
\be
[\tB_{\alpha a},\tX_\beta]
=\frac 1 \omega A^{-1}[B_{\alpha a},X_\beta]=i\eta_{\alpha\beta}
\frac 1 \omega A^{-1} X_a
=i\eta_{\alpha\beta}\frac 1 \omega A^{-1}B\tX_a
\ee
and
\be
[\tB_{\alpha a},\tX_b]
=\frac 1 \omega B^{-1}[B_{\alpha a},X_b]=-i\eta_{ab}
\frac 1 \omega B^{-1} X_\alpha
=-i\eta_{ab}\frac 1 \omega B^{-1}A\tX_a
\ee
Defining the following matrix:
\be 
C= A^{-1}B,
\ee
the previous commutators are given by:
\be\label{new}
[\tB_{\alpha a},\tX_\beta]=i\eta_{\alpha\beta}\frac 1\omega C \tX_a,~~~[\tB_{\alpha a},\tX_b]=-i\eta_{ab}\frac 1\omega C^{-1} \tX_\alpha
\ee 
We are now in the position to prove the no-go theorem:\\

{\it Any  $k$-contraction of the extension  of the Poincar\'e algebra with a Lorentz vector $Z_\mu$ cannot give rise to the Galilei algebra with  extension verifying
(\ref{eq:9}). }\\

In fact, the commutator $[\tB_{\alpha a}, \tP_\beta]=i\eta_{\alpha\beta}\tP_a$ of  \eq{9} requires that its r.h.s. should not contain a component along $\tZ_a$, which amounts to require the following condition for the matrix element $C_{12}$ of the matrix $C$:
\be
\lim_{\omega\to\infty} \frac 1\omega C_{12} =0,~~~\rightarrow~~~C_{12} =b(\omega)\omega^{1-\alpha}\,~~~{\rm with}~~~\lim_{\omega\to\infty} b(\omega) =b\label{eq:20}
\ee
with $b$ finite and $\alpha>0$. Then, using the second of eqs. \eq{9}:
\be
\lim_{\omega\to\infty} \frac 1\omega (C^{-1})_{12} =-\lim_{\omega\to\infty} \frac 1{\omega\Delta} C_{12}
={\rm constant}\not= 0
\ee
where $\Delta$ is the determinant of $C$.
From the result in \eq{20} we obtain:
\be 
\Delta= c(\omega) \omega^{-\alpha},~~~\lim_{\omega\to\infty} c(\omega)=c\not=0
\ee
Then, using again the first equation \eq{9}, implying that in the limit $C_{11}/\omega$ goes to a constant, we get:
\be
\frac{ (C^{-1})_{22}}\omega=\frac{ C_{11}}{\omega\Delta}\to \omega^\alpha
\ee
Therefore, requiring that both eqs. \eq{9} are satisfied we obtain a non convergent limit, and the no-go theorem is proved.

In order to proceed, we will  look for a 
 contraction as close as possible to the results of  eqs. \eq{9}. To this end, it is enough to take $\alpha=0$ in the discussion made in the proof of the no-go theorem. This amounts to take, in the limit $\omega\to\infty$
\be
\frac 1\omega C_{11}=a(\omega), ~~~\frac 1\omega C_{12}=b(\omega)
\ee
with the determinant $\Delta$ and the parameters  $a(\omega)$ and $ b(\omega)$ all going to  a finite limit. It follows that  the matrices $C$ and $C^{-1}$ should be of the following form
\be
C=\left(\begin{array}{cc}a\omega & b\omega \\\frac c\omega & \frac d\omega\end{array}\right),~~~
C^{-1}=\frac 1\Delta\left(\begin{array}{cc}\frac d\omega & -b\omega \\ -\frac c\omega & a\omega\end{array}\right),~~\Delta=ad-bc
\ee
Notice that the terms proportional to $1/\omega$ come from the requirement to get $\Delta$ finite.
Therefore, in the limit $\omega\to\infty$ we have
\be
\lim_{\omega\to\infty}\frac 1\omega  C=\left(\begin{array}{cc}a &b \\ 0 & 0\end{array}\right)
\ee
\be
\lim_{\omega\to\infty}\frac 1\omega C^{-1}=\frac 1\Delta\left(\begin{array}{cc}0 &-b \\ 0 &a\end{array}\right),
\ee
The algebra (\ref{new}) becomes
\be
[\tB_{\alpha a}, \tP_\beta]=i\eta_{\alpha\beta}\left(a\tP_a +b \tZ_a\right),~~~
[\tB_{\alpha a}, \tZ_\beta]=0 \label{eq:37}
\ee
\be
[\tB_{\alpha a}, \tP_b]=i\eta_{ab}\frac b\Delta \tZ_\alpha ,~~~[\tB_{\alpha a}, \tZ_b]=- i\eta_{ab}\frac a\Delta  \tZ_\alpha\label{eq:38}
\ee
This algebra depends on three independent parameters, $a,b,\Delta$, However, assuming that  they are different from zero, it is easy to show that they can be  eliminated  by a rescaling of some of the generators without changing their commutators with the other group generators. This is equivalent to take all the parameters equal to one.
By doing that it follows
\be
[\tB_{\alpha a}, \tP_\beta]=i\eta_{\alpha\beta}\left(\tP_a +\tZ_a\right),~~~
[\tB_{\alpha a}, \tZ_\beta]=0 \label{eq:40}
\ee
\be
[\tB_{\alpha a}, \tP_b]=i\eta_{ab} \tZ_\alpha ,~~~[\tB_{\alpha a}, \tZ_b]=- i\eta_{ab} \tZ_\alpha\label{eq:41}
\ee
Notice that it would not help to define a new momentum
\be
\tP_a +\tZ_a
\ee
since this commute with the boosts.

We will not discuss here the case when $a$ or $b$ are vanishing since in these cases we do not get interesting algebras.

To summarise, what goes wrong here for obtaining the eqs. \eq{9} is the presence of the term $\tZ_a$ in the commutator of the boost with $\tP_\alpha$, and there is no way to eliminate this term within this extended algebra. A clear possibility would be to mix the boosts with a new term such to eliminate the $\tZ_a$ contribution. By covariance we would need to add a term transforming as $\tB_{\alpha a}$ under the Lorentz group in $k$ dimensions and the rotation group in $D+k-1$ dimensions. Therefore, from the point of view of the Lorentz group in $D+1$ dimension we need an  
 antisymmetric tensor of rank two.

\section{A further extension of the algebra}

As we have seen in the previous Section, in order to overcome the problem of the no-go theorem, we would need a further extension of the Poincar\'e algebra by an antisymmetric two-tensor $Z_{\mu\nu}$, besides the vector charges $Z_\mu$.

 A natural framework will be to consider the vector space of the whole Poincar\'e Lie algebra  and its dual. The generators will be $(P_\mu, M_{\mu\nu})\oplus 
(P_{\mu}^*=Z_\mu, M_{\mu\nu}^*=Z_{\mu\nu})$. In this case the technique of the Appendix gives the same result of the double extension procedure.  The resulting algebra is called the coadjoint Poincar\'e algebra and it has a non degenerate metric \cite{medina,Figueroa-OFarrill:1995opp,Matulich:2019cdo}.

%
%
We have,
 \bea
  \left[M_{\mu\nu},Z_{\rho}\right]&=&i(\eta_{\mu\rho}Z_{\nu}-\eta_{\nu\rho}Z_{\mu}),
\nn\\
 \left[M_{\mu\nu},Z_{\rho\sigma}\right]&=&i(\eta_{\mu\rho}Z_{\nu\sigma}+\eta_{\nu\sigma}Z_{\mu\rho}
 -\eta_{\mu\sigma}Z_{\nu\rho}
  -\eta_{\nu\rho}Z_{\mu\sigma})\nn\\
    \left[Z_{\mu\nu},P_{\rho}\right]&=&i(\eta_{\mu\rho}Z_{\nu}-\eta_{\nu\rho}Z_{\mu}),
\nn\\
  \left[Z_{\mu},P_{\nu}\right]&=&0 ,~~~  \left[Z_{\mu},Z_{\nu}\right]=0,~~~
,~~~ \left[Z_{\mu\nu},Z_{\rho}\right]=0
  \label{eq:1.18}
\eea
If we consider the ideal generated by $Z_{\mu\nu}$ and compute the corresponding quotient we reproduce the Poincar\'e algebra extended by a vector charge $Z_\mu$ introduced in Section 3.
 
 Mixing  together $B_{\alpha a}$ and $Z_{\alpha a}$  makes possible to cancel the term in $\tZ_a$ present in the commutator $[\tB_{\alpha a},\tP_\beta]$, that was plaguing us
 in the previous Section. 

The following definitions satisfy our request:
\be
P_\alpha=\tP_\alpha+\frac 12\omega^2\tZ_\alpha,~~~Z_\alpha=\omega\tZ_\alpha,
~~~P_a=\omega\tP_a,
\ee
with the inversion relations:
\be
\tP_\alpha= P_\alpha-\frac 1 {2}\omega Z_\alpha,~~~\tZ_\alpha=\frac 1{ \omega}Z_\alpha,~~~\tP_a=\frac 1 \omega P_a,
\ee
Furthermore, we define
\be
\tB_{\alpha a}=\frac 1\omega B_{\alpha a} + \frac 12 Z_{\alpha a}
\ee

With these definitions we have 
\bea
[\tB_{\alpha a}, \tP_\beta]&=&i\eta_{\alpha\beta}\tP_a,~~~ [\tB_{\alpha a}, \tP_b]=- i\eta_{ab}\tZ_\alpha\label{eq:53}
\eea 
\be
[\tB_{\alpha a}, \tZ_\beta]=0,\label{eq:48}
\ee
therefore the  eqs. \eq{53} reproduce the wanted result given in \eq{9}.

Furthermore, assuming
\be
\tZ_a= \omega^{-\eta} Z_a,~~~\tZ_{\alpha a}=\omega^{\xi} Z_{\alpha a}, 
\ee
we get
\be
[\tZ_{\alpha a}, \tZ_{\beta b}]=[\tZ_{\alpha a}, \tZ_{\beta b}]=0\label{eq:50}\ee
\be
[\tB_{\alpha a}, \tZ_b]=
-i\eta_{ab}\,\omega^{-\eta} \tZ_\alpha,~~~[\tZ_{\alpha a},\tP_\beta]=i\omega^{\xi+\eta}\eta_{\alpha\beta}  \tZ_a~~~
[\tZ_{\alpha a},\tP_b]=-i\eta_{ab}\omega^{\xi}\tZ_\alpha\label{eq:51}
\ee
implying 
\be
\eta\ge 0,~~~\xi+\eta\le 0,~~~\xi\le 0\label{eq:52}
\ee
As for the generators $M_{\alpha\beta}$ and $M_{ab}$, requiring that the contracted generators span respectively a Lorentz and an euclidean algebra, implies  no mixing with $Z_{\alpha\beta}$ and $Z_{ab}$ is  introduced. Therefore:
\be
\tM_{\alpha\beta}=M_{\alpha\beta},~~~\tM_{ab}=M_{ab}\ee
whereas we will assume
\be
 \tZ_{\alpha\beta}=\omega^{\tau}Z_{\alpha\beta},~~~\tZ_{ab}=\omega^{\sigma}Z_{ab}
\ee
Notice  that  the commutator of 
$\tM_{\alpha\beta}$ and $\tM_{ab}$ with any other contracted generator $\tX$ is equal to the contraction of  the commutator of $M_{\alpha\beta}$ and $M_{ab}$   with $X$.

It follows
\be
[\tB_{\alpha a},\tB_{\beta b}]=i\eta_{\alpha\beta} \omega^{-\sigma-1} \tZ_{ab}+i\eta_{ab}\omega^{-\tau-1} \tZ_{\alpha\beta}\label{eq:54}
\ee
\be
[\tZ_{\alpha\beta}, \tP_\gamma]=i \omega^{\tau+1}(\eta_{\alpha\gamma}\tZ_\beta-\eta_{\beta\gamma}\tZ_\alpha)\label{eq:55}
\ee
\be
[\tZ_{ab}, \tP_c]=i \omega^{\sigma-1{\color{red}+\eta}}(\eta_{ac}\tZ_b-\eta_{bc}\tZ_a)\label{eq:56}
\ee
\be
[\tZ_{\alpha\beta}, \tB_{\gamma a}]=i\omega^{\tau-\xi-1} (\eta_{\alpha\gamma}\tZ_{\beta a}-\eta_{\beta\gamma}\tZ_{\alpha a})\label{eq:57}
\ee
\be
[\tZ_{ab}, \tB_{\alpha c}]=i\omega^{\sigma-\xi-1} (\eta_{ac}\tZ_{\alpha b}-\eta_{bc}\tZ_{\alpha a})\label{eq:58}
\ee
From these expressions and the conditions in \eq{52} we get the restrictions on the parameters
\be
\tau =-1,~~~-1\le\sigma\le1-\eta,~~~-2\le\xi\le 0,, ~~\eta\ge 0,~~~\xi+\eta\le 0,~~\sigma-\xi -1\le 0\label{eq:60}\ee
from which we get $0\le \eta\le 2$.

Therefore we get as many contractions giving rise to the Galilei algebra 
 \eq{53} as many are the sets of exponents satisfying the inequalities in \eq{60}.

\begin{figure}[h]
\begin{center}
   \includegraphics[width=4in]{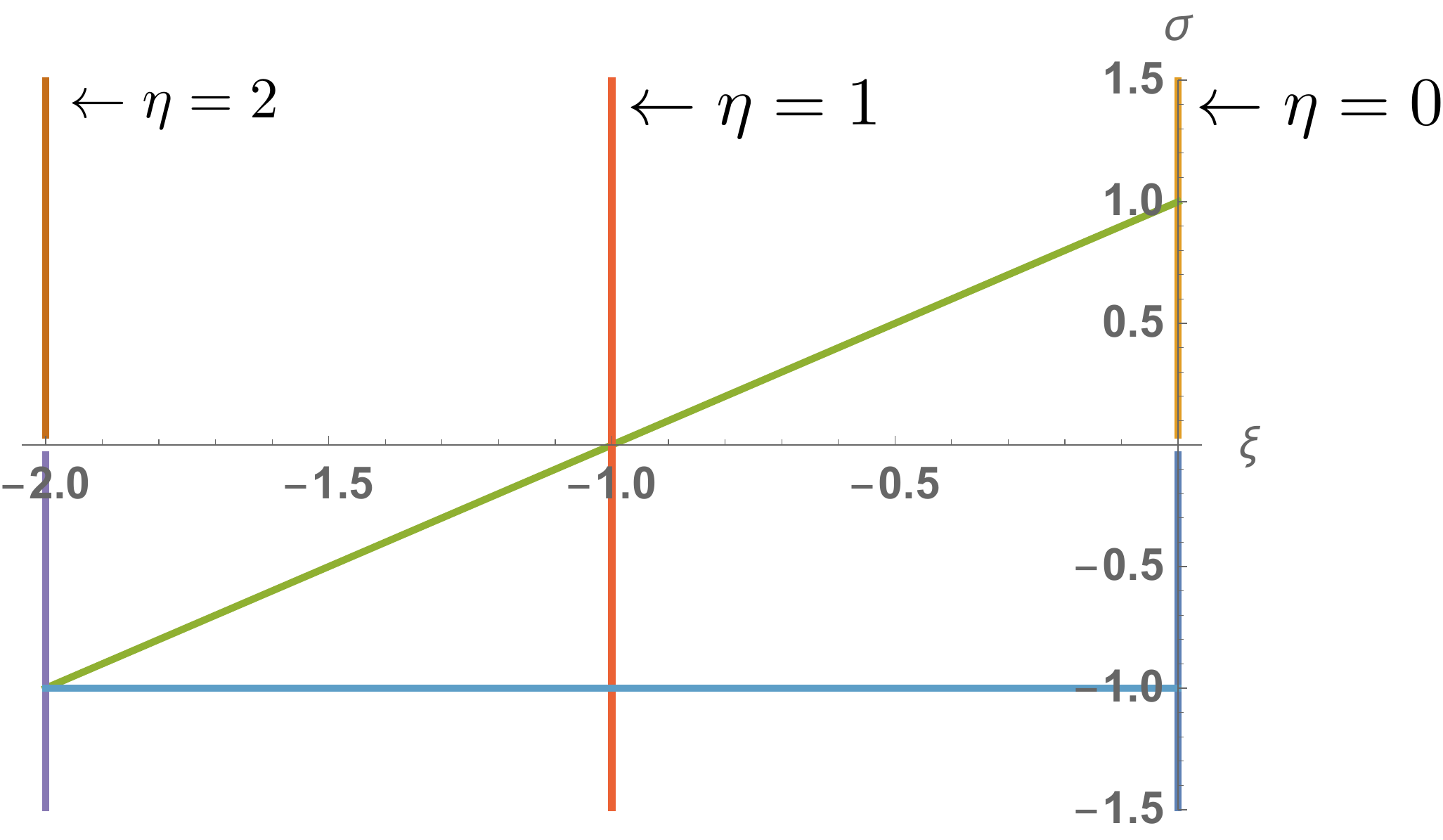} 
  \caption{The allowed region for the contraction is the inside and the borders of the triangle in the figure. The three vertical lines are the limiting ones for the cases $\eta=0,1,2$}
  \label{fig:1}
  \end{center}
  \end{figure}

Let us now consider the various possibilities for $\eta$: \\\\
\noindent
 $\eta=0$. 
Then, the inequalities in \eq{60} become:
\be
\tau=-1,~~~ -2\le\xi\le 0,~~~  -1\le\sigma\le 1, ~~~ \sigma-\xi -1\le 0
\ee
These conditions give rise to the triangular region in the plane $(\xi,\sigma)$ given in Fig. \ref{fig:1}.
  
Therefore the solutions for integer values of $\xi$ and $\sigma$ are::\\
\noindent
1a): $\tau=-1,,\xi=-2, \sigma =-1$;\\
\noindent
2a) : $\tau=-1,,\xi=-1, \sigma =-1$;\\
\noindent
3a) : $\tau=-1,,\xi=-1, \sigma =0$;\\
\noindent
4a) : $\tau=-1,,\xi=0, \sigma =-1$;\\
\noindent
5a) : $\tau=-1,,\xi=0, \sigma =0$;\\
\noindent
6a) : $\tau=-1,,\xi=0, \sigma =+1$;\\
\noindent
$\eta=+1$. In this case the inequalities \eq{60} become:
\be
\tau =-1,~~~-1\le\sigma\le 0,~~~-2\le\xi\le -1,~~\sigma-\xi -1\le 0
\ee
The solutions are again in the triangle of Fig. {\ref{fig:1}, but this time delimited from the right by the line $\xi=-1$. Therefore we have three solutions:\\ 
\noindent
1b) : $\tau =-1, ~~~\eta =1,~~~, \sigma =-1,~~~ \xi =-1$,\\
\noindent
2b):  $\tau =-1, ~~~\eta =1,~~~, \sigma =0, \xi =-1$\\
\noindent
3b):  $\tau =-1, ~~~\eta =1,~~~, \sigma =-1, \xi =-2$\\\\
\noindent
$\eta=+2$.  }From the inequalities \eq{60}:\\
\noindent
1c): $\tau =-1, ~~~\eta =2,~~~, \sigma =-1, \xi =-2$\\
\noindent
We get a single solution corresponding to the left vertex of the triangle in  Fig. {\ref{fig:1}.

Summarising, the commutators \eq{53}, \eq{48} and \eq{50} are common to all the contractions, whereas the ones depending on the solutions of inequalities \eq{60}, that is:

\be
[\tB_{\alpha a}, \tZ_b]=
-i\eta_{ab}\,\omega^{-\eta} \tZ_\alpha,~~~[\tZ_{\alpha a},\tP_\beta]=i\omega^{\xi+\eta}\eta_{\alpha\beta}  \tZ_a~~~
[\tZ_{\alpha a},\tP_b]=-i\eta_{ab}\omega^{\xi}\tZ_\alpha\label{eq:51}
\ee
\be
[\tB_{\alpha a},\tB_{\beta b}]=i\eta_{\alpha\beta} \omega^{-\sigma-1} \tZ_{ab}+i\eta_{ab}\omega^{-\tau-1} \tZ_{\alpha\beta}\label{eq:54}
\ee
\be
[\tZ_{\alpha\beta}, \tP_\gamma]=i \omega^{\tau+1}(\eta_{\alpha\gamma}\tZ_\beta-\eta_{\beta\gamma}\tZ_\alpha)\label{eq:55}
\ee
\be
[\tZ_{ab}, \tP_c]=i \omega^{\sigma-1+\eta}(\eta_{ac}\tZ_b-\eta_{bc}\tZ_a)\label{eq:56}
\ee
\be
[\tZ_{\alpha\beta}, \tB_{\gamma a}]=i\omega^{\tau-\xi-1} (\eta_{\alpha\gamma}\tZ_{\beta a}-\eta_{\beta\gamma}\tZ_{\alpha a})\label{eq:57}
\ee
\be
[\tZ_{ab}, \tB_{\alpha c}]=i\omega^{\sigma-\xi-1} (\eta_{ac}\tZ_{\alpha b}-\eta_{bc}\tZ_{\alpha a})\label{eq:58}
\ee
are listed in Table \ref{table:1}

\begin{table}[h]
\caption{The commutators corresponding to the different solutions of  the inequalities \eq{60}. The notations $xa)$, $xb)$ and $xc)$ refer to the cases $\eta=0,1,2$ respectively. The square brackets in the lower indices refer to the antisymmetrization.}
\begin{center}
\begin{tabular}{||c||c|c|c|c|c|c|c|c||}
\hline
 Sol.&~$[\tB_{\alpha a}, \tZ_b]$~&~$[\tZ_{\alpha a},\tP_\beta]$~ &~$[\tZ_{\alpha a},\tP_b]$~&~$[\tB_{\alpha a},\tB_{\beta b}] $ &~ $[\tZ_{\alpha\beta}, \tP_\gamma] $ ~&$[\tZ_{ab}, \tP_c]$ &~$[\tZ_{\alpha\beta}, \tB_{\gamma a}]$~ &~$[\tZ_{ab}, \tB_{\alpha c}]$~\\
\hline\hline
1a)&~$-i\eta_{ab}\tZ_\alpha$~ & 0 & 0 & $i\eta_{\alpha\beta} \tZ_{ab}+i\eta_{ab} \tZ_{\alpha\beta}$ & $i\eta_{[\alpha\gamma}\tZ_{\beta]}$ & $0$ & $i \eta_{[\alpha\gamma}\tZ_{\beta]\ a}$ &$i \eta_{[ac}\tZ_{\alpha b]}$ \\
\hline
2a) &~$-i\eta_{ab}\tZ_\alpha$~ & 0 & 0 & $i\eta_{\alpha\beta} \tZ_{ab}+i\eta_{ab} \tZ_{\alpha\beta}$ & $i\eta_{[\alpha\gamma}\tZ_{\beta]}$ & $0$ & $0$ &$0$\\
\hline
3a) &~$-i\eta_{ab}\tZ_\alpha$~ & 0 & 0 & $i\eta_{ab} \tZ_{\alpha\beta}$ & $i\eta_{[\alpha\gamma}\tZ_{\beta]}$ & $0$ & $0$ &$i \eta_{[ac}\tZ_{\alpha b]}$\\
\hline
4a) &~$-i\eta_{ab}\tZ_\alpha$~ &$ i\eta_{\alpha\beta} \tZ_a$ & $-i\eta_{ab}\tZ_\alpha$ & $i\eta_{\alpha\beta} \tZ_{ab}+i\eta_{ab} \tZ_{\alpha\beta}$ & $i\eta_{[\alpha\gamma}\tZ_{\beta]}$ & $0$ & $0$ &$0$\\
\hline
5a) &~$-i\eta_{ab}\tZ_\alpha$~ &$ i\eta_{\alpha\beta} \tZ_a$ & $-i\eta_{ab}\tZ_\alpha$ & $i\eta_{ab} \tZ_{\alpha\beta}$ & $i\eta_{[\alpha\gamma}\tZ_{\beta]}$ & $0$ & $0$ &$0$\\
\hline
6a) &~$-i\eta_{ab}\tZ_\alpha$~ &$ i\eta_{\alpha\beta} \tZ_a$ & $-i\eta_{ab}\tZ_\alpha$ & $i\eta_{ab} \tZ_{\alpha\beta}$ & $i\eta_{[\alpha\gamma}\tZ_{\beta]}$ & $i \eta_{[ac}\tZ_{b]}$ & $0$ &$i \eta_{[ac}\tZ_{\alpha b]}$\\
\hline\hline
1b) &~$0$~ &$ i\eta_{\alpha\beta} \tZ_a$ & $0$ & $i\eta_{\alpha\beta} \tZ_{ab}+i\eta_{ab} \tZ_{\alpha\beta}$ & $i\eta_{[\alpha\gamma}\tZ_{\beta]}$ & $0$ & $0$ &$0$\\
\hline
2b) &~$0$~ &$ i\eta_{\alpha\beta} \tZ_a$ & $0$ & $i\eta_{ab} \tZ_{\alpha\beta}$ & $i\eta_{[\alpha\gamma}\tZ_{\beta]}$ & $i \eta_{[ac}\tZ_{b]}$ & $0$ &$i \eta_{[ac}\tZ_{\alpha b]}$\\
\hline
3b) &~$0$~ &$0$ & $0$ & $i\eta_{\alpha\beta} \tZ_{ab}+i\eta_{ab} \tZ_{\alpha\beta}$ & $i\eta_{[\alpha\gamma}\tZ_{\beta]}$ & $0$ & $i \eta_{[\alpha\gamma}\tZ_{\beta]\ a}$ &$i \eta_{[ac}\tZ_{\alpha b]}$\\
\hline\hline
1c) &~$0$~ &$ i\eta_{\alpha\beta} \tZ_a$ & $0$ & $i\eta_{\alpha\beta} \tZ_{ab}+i\eta_{ab} \tZ_{\alpha\beta}$ & $i\eta_{[\alpha\gamma}\tZ_{\beta]}$ & $i \eta_{[ac}\tZ_{b]}$ & $i \eta_{[\alpha\gamma}\tZ_{\beta]\ a}$ &$i \eta_{[ac}\tZ_{\alpha b]}$\\
\hline\hline
\end{tabular}
\end{center}
\label{table:1}
\end{table}

Let us discuss the results. First of all we notice that in order to reproduce the vanishing of the commutator of the boosts among themselves for $k=1$, we need that the term in $\tZ_{ab}$ is not present. From the \eq{54} we see that this happens only for $\sigma=0,1$ or, from Table \ref{table:1} in the cases 3a), 5a), 6a)  and 2b). In these cases, since the antisymmetric tensor $\tZ_{\alpha\beta}$ vanishes for $k=1$ we obtain  the ordinary  Galilei algebra. In the other cases we obtain a generalization of the Galilei group with non-commuting boosts also for $k=1$. In this last case the commutator is proportional to the antisymmetric tensor $\tZ_{ab}$.
Then, we can say that enlarging the Poincar\'e algebra with a vector and an antisymmetric tensor, there is no way of getting commuting boosts for $k>1$.

A particular interesting case is the 2b). In fact, let us consider the following decomposition of the full Lie algebra, ${\cal G}$
, corresponding to this case:
\be
{\cal G}={\cal A}\oplus{\cal I},~~~{\cal A}=(\tM_{\alpha\beta}, \tM_{a b}, \tP_\alpha, \tP_a, \tM_{\alpha a}, \tZ_\alpha),~~~{\cal I}=(\tZ_{ab}, \tZ_a, \tZ_{\alpha a})
\ee
From Table \ref{table:1} it follows that ${\cal A}$ is a subalgebra of  ${\cal G}$ and ${\cal I}$ is an ideal. In fact:
\be
[{\cal A}, {\cal A}]={\cal A},~~~[{\cal A}, {\cal I}]={\cal I},~~~[{\cal I}, {\cal I}]={\cal I}.
\ee
By taking the quotient ${\cal G}/{\cal I}$ we have
the  (stringy) p-brane Galilei algebra \cite{Brugues:2004an,Brugues:2006yd}
 used in the study of non-relativistic strings \cite{Gomis:2000bd,Danielsson:2000gi}  and the stringy Newton Cartan gravity
  \cite{Andringa:2012uz,Bergshoeff:2018vfn}.
  
 The generators
of this algebra are the longitudinal and traverse translations $\tP_\alpha,\tP_a$, longitudinal and transverse rotations
 $\tM_{\alpha\beta},  \tM_{ab}$,  generalized NR boots $\tB_{\alpha a}$,
and non-central charges $ \tZ_\alpha,
\tZ_{\alpha\beta}$ where $\alpha,\beta={0,1\cdots k-1}; a,b={k,\cdots D}$.
Here we write the commutations relations that involve the non-central charges
\bea\label{stringy}
[\tB_{\alpha a}, \tP_\beta]&=&i\eta_{\alpha\beta} b\tP_a,~~~[\tB_{\alpha a},\tB_{\beta b}]=i\eta_{ab} \tilde \tZ_{\alpha\beta}
\\
 \left[\tM_{\alpha\beta},\tZ_{\gamma}\right]&=&i(\eta_{\alpha\gamma}\tZ_{\beta}-\eta_{\beta\gamma}\tZ_{\alpha}),
\\
 \left[\tM_{\alpha\beta},\tZ_{\gamma\delta}\right]&=&i(\eta_{\mu\rho}\tM_{\nu\sigma}+\eta_{\nu\sigma}\tM_{\mu\rho}
 -\eta_{\mu\sigma}\tM_{\nu\rho}
  -\eta_{\nu\rho}\tM_{\mu\sigma}),
\\
 \left[\tZ_{\alpha\beta},\tP_\gamma\right]&= &i\eta_{\alpha\gamma}\tZ_\beta- 
 i\eta_{\beta\gamma}\tZ_\alpha.
 \eea

Also in other cases we have ideals among the $Z's$. For instance, with ${\cal I} =(\tZ_a, \tZ_{\alpha a})$, there are the cases 1b)
 and 1c). With ${\cal I}=(\tZ_{\alpha a}, \tZ_{ab})$ the case 3a). In the cases 2a), 5a) and 3b) we have ${\cal I}= \tZ_{\alpha a}$,  ${\cal I}= \tZ_{ab}$,  ${\cal I}= \tZ_{\alpha a}$,  ${\cal I}= \tZ_a$ respectively. Finally,  in the cases 1a), 4a) 6a)and 3b) there are no ideals among the $Z's$.
 
\section{Conclusions and outlook}
In this paper we have constructed extensions of the k-contracted relativistic Poincar\'e algebras 
with extra generators with the request of having a generalization of the commutation relations of the Bargmann algebra.
We have proved the need to consider as the relativistic starting point  the coadjoint Poincar\'e algebra with generators $P_{\mu}, M_{\mu\nu}$
$Z_{\mu}, Z_{\mu\nu}$. We have found a series of non-relativistic algebras verifying this property.
In one of the cases we have  found that there is an ideal such that the quotient of the whole algebra with the ideal gives the stringy (p-brane) Galilei algebra 
\cite{Brugues:2006yd,  Brugues:2004an}
that has been used in the literature as a symmetry of non-relativistic strings 
\cite{Gomis:2000bd,Danielsson:2000gi}
and stringy Newton-Cartan gravity \cite{Andringa:2012uz,Bergshoeff:2018vfn,Gomis:2019zyu,Aviles:2019xed,Gallegos:2019icg}.

There are several points that  could be further studied. For example, are there other interesting contractions of the coadjoint Poincar\'e algebra that do no respect the eqs.(\ref{eq:9})?
What are the dynamical realisations that  can be constructed from these algebras using the method of non-linear realisations? Can we construct non-relativistic gravities associated to the algebras we have found?

 \section*{Acknowledgments}
 We acknowledge discussions with Adolfo Azcarraga, Eric Bergshoeff,  Diederik Roest,  Patricio Salgado-Rebolledo and Tonnis ter Veldhuis.
 JG acknowledges the hospitality and support of the Van Swinderem Institute where this work was finished.
 JG also has been supported in part by MINECO FPA2016-76005-C2-1-P and Consolider CPAN,
and by the Spanish government (MINECO/FEDER) under project MDM-2014-0369 of
ICCUB (Unidad de Excelencia Mar\`a de Maeztu).

\section{Appendix: Double Extensions}

We will consider  possible extensions of a Lie algebra by the addition of vector spaces isomorphic to an ideal of the algebra  or to the algebra itself.

This procedure is inspired by an extension that in the literature is known as
 {\it double extension} \cite{medina,Figueroa-OFarrill:1995opp}, (see also \cite{Matulich:2019cdo}).  Usually this extension is made  by means of one or more central or non-central extensions which have associated a non-degenerate invariant symmetric bilinear form \cite{medina}, but we will do not need here the existence of this form. As a consequence,  we will define the double extension with no reference to a particular basis in the Lie algebra,

We start considering a Lie algebra with the  structure of a vector space
 \be
 \cG=\cM\oplus\cP
 \ee
and with the following commutation relations:
\be
[\cM,\cM]\in\cM.~~~[\cM,\cP]\in \cP,~~~[\cP,\cP]\in\cP
\ee
meaning that $\cP$ is an ideal of $\cG$.
Then, suppose that we are given a vector space isomorphic to $\cP$, say, $\cP^*$. Then, we can establish  a one-to-one correspondence among the  two vector spaces $\cP$ and $\cP^*$:
\be
O\rightarrow O^*,~~~O\in\cP,~~O^*\in \cP^*
\ee
We can extend the Lie algebra $\cG$ to $\cG\oplus\cP^*=\cM\oplus\cP\oplus\cP^*$, through the following definitions:
\be
[\cM,\cP^*]=[\cM,\cP]^*,~~~[\cP,\cP^*]=[\cP,\cP]^*,~~~[\cP^*,\cP^*]=0
\ee
with
\be
[\cM,\cP^*]=-[\cP^*,\cM]~~~[A,B^*]=-[B^*,A]~~~A,B\in \cP
\ee
The following property holds
\be
[\cP^*,\cM]=-[\cM,\cP^*]=-[\cM,\cP]^*=[\cP,\cM]^*
\ee
and
\be
[A^*,B]=[A,B^*],~~~A,B\in \cP
\ee
as follows from
\be
[A^*,B]=-[B,A^*]=-[B,A]^*=[A,B]^*=[A,B^*]
\ee
 Then, it is trivial to check the validity of the Jacobi identity. For instance, consider the following triple commutator
 \be
 [\cM,[\cM,\cP^*]]= [\cM,[\cM,\cP]^*]= [\cM,[\cM,\cP]]^*
 \ee
 Analogously we get
 \be
 [\cM,[\cP^*,\cM]]=  [\cM,[\cP,\cM]]^* , ~~~ [\cP^*,[\cM,\cM]]=  [\cP,[\cM,\cM]]^*
 \ee
 Summing together these three relations we see that the Jacobi identity of the enlarged algebra follows from the Jacobi identity of the original algebra.
 
In particular, in the main text,  we have applied this procedure to the Poincar\'e algebra extended with the vector charges $Z_\mu$.


Consider now a Lie algebra  ${\cal G}$ . Then, suppose to have another vector space, ${\cal G}^*$ isomorphic to the vector space ${\cal G}$. This means that we can establish a linear  one-to-one correspondence among the elements of ${\cal G}$ and the elements of ${\cal G}^*$:
\be
A\rightarrow A^*,~~ A\in {\cal G},~~ A^*\in{\cal G}^*
\ee
We can extend the Lie structure to  the entire space ${\cal G}\oplus{\cal G}^*$ by defining the  following parenthesis:
\be
[A,B^*]=[A,B]^*,~~~[B^*,A]=-[A,B^*],~~~[A^*,B^*]=0
\ee
Notice that
\be
[A^*,B]=-[B,A^*]=-[B,A]^*=[A,B]^*=[A,B^*]
\ee
that is
\be
[A^*,B]=[A,B^*]\label{eq:75}
\ee
 These parenthesis are antisymmetric by construction and satisfy the Jacobi identity. In fact, consider the triple commutator
\be
[A,[B,C^*]]=[A,[B,C]^*]=[A,[B,C]]^*
\ee
Then, using \eq{75} and the previous relation  we have
\be
[B,[C^*,A]]=[B,[C,A]]^*,~~~[C^*,[A,B]]=[C,[A,B]]^*
\ee
 By linearity, the sum of these triple commutators is the star of the Jacobi identity for three elements of ${\cal G}$. Therefore two elements of ${\cal G}$ and one of ${\cal G}^*$ satisfy the  Jacobi identity. The same holds for the case 
  of one element of ${\cal G}$ and two elements of ${\cal G}^*$, and  for the case of  three elements of ${\cal G}^*$.  Therefore this extension of the Lie algebra  ${\cal G}$  to ${\cal G}\oplus{\cal G}^*$ is a Lie algebra.

 In the main text we have applied this procedure to extend the Poincar\'e algebra with the vector space generated by the charges $(Z_\mu,Z_{\mu\nu})$, which form a vector space isomorphic to the vector space of the generators $(P_\mu,M_{\mu\nu})$.

%
%

\end{document}